# Understanding The Top 10 OWASP Vulnerabilities

Matthew Bach-Nutman
*Bournemouth University*
Bournemouth, United Kingdom
s5085361@bournemouth.ac.uk

*Abstract*

Understanding the common vulnerabilities in web applications help businesses be better prepared in protecting their data against such attacks. With the knowledge gained from research users and developers can be better equipped to deal with the most common attacks and form solutions to prevent future attacks against their web applications. Vulnerabilities exist in many forms within modern web applications which can be easily mitigated with investment of time and research.

## I. Introduction

With the rise of cyber-attacks in the past decade [1], websites are vastly targeted and are being exploited leaving businesses with substantial losses [1]. Web development projects are constructed without security in mind which can lead to loss of assets, disruption of service, exposure of sensitive data or system compromise. This report will aim to provide readers with a better understanding about the top 10 Open Web Application Security Project (OWASP) vulnerabilities, which actions should be taken to mitigate these vulnerabilities, the impact on businesses who suffer from an attack and reaching the conclusion whether security should be taken seriously.

## II. Defining the problem

### A. What are Injections?

Injections attacks [2] refer to a broad class of attack vectors [3]. Injection attacks occur when attackers find weaknesses in input validation and enter untrusted inputs which then gets processed by the web application and alters the intended execution of the backend processing.

- SQL Injection [2] attacks occur when user input is not correctly sanitized and have amongst the biggest impact on the internet. When attackers find a SQL injection, in most cases they can retrieve data from the database and in some cases lead to remote code execution on the target system. SQL injections attacks come in many forms such as error-based attacks, union-based attacks, blind attacks and out-of-band attacks [4]. Error-based attacks are when injections occur and produce SQL errors which can disclose the data within the database. Union-based attacks are when injections continue a previous SQL statement to obtain the data from the database using the UNION clause which joins a secondary statement to the existing statement. Blind-based attacks occur when the injections return no output or errors, so an attacker must get creative in order to determine the output. Out-of-band attacks are when the database does not return any output but allows an attacker to feed the injection output to a source where they can read such as their own HTTP server. Mitigation of these kind of these attacks requires developers to properly validate user input or use prepared statements [4] when communicating with the database from a web application.

- Code injections [2] occur when attackers exploit input validation flaws in websites and software which allows attackers to execute arbitrary commands on the target host. Code is injected in the language of the targeted application and is executed by the server-side interpreter for that language [5]. There are multiple backend languages and enumeration must occur for attackers to establish which language is being used. Some of the common languages are PHP, Java, Python and Ruby. Mitigation against code injection attacks include validating user input, removing system functions from the application, treating all data as untrusted [5] and ensuring that the application code has been checked before use.

- The Lightweight Directory Application Protocol (LDAP) is a protocol which is used to distribute lists of information organized into directory information trees which are stored within the LDAP database [6].

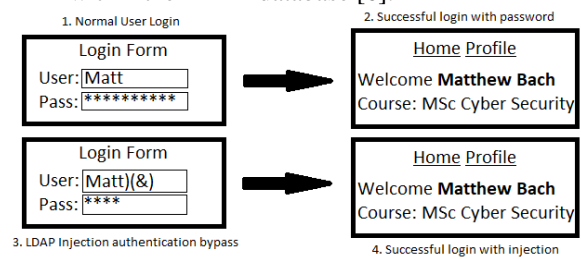

*Figure 1 - LDAP Authentication Bypass Example*

LDAP injections [2] occur when attackers manipulate search capabilities of the protocol which is like SQL injections [7]. Figure 1 shows a simple authentication bypass on a login form utilizing LDAP injection. Mitigation against LDAP injections include filtering of special characters at the application layer [7] so users cannot inject queries into the web application.

- XPATH injection [2] attacks occur when websites construct an XPath query for XML data from user supplied information [8]. The attack is carried out when malicious user inputs are injected into the site using query strings which can lead to unauthorized access or reveal sensitive information. There are 2 types of attacks in XPATH injections which are Boolean-based [8] attacks where attackers can inject queries that lead to true or false server responses or XML crawling [8] which allows attackers to extract sensitive information by injecting specific queries within an XML document which is processed by the website. Mitigation against these types of attacks are like SQL injections and require sanitizing user input or using precompiled statements which pass user input as parameters instead of expressions [8].

- Host headers are used when many web applications are hosted on the same IP address and are used to specify which application should process a HTTP request [9]. Host header injection [2] attacks occur when attackers injects a malicious host into the HTTP host header which allows attackers to control injections on the first virtual host which can lead to arbitrary code execution in the context of the target website [9]. A common attack which host header injections can lead to is web-cache poisoning which is where attackers inject content which gets stored into the web cache and is rendered by anyone who requests the resource [9]. Another common attack is password reset poisoning which can occur from host header injections when a web application uses host headers while generating password reset links, attackers can inject host headers to poison the reset link which gets sent to victims allowing the attacker to gain access to the password reset token and authenticate as the victim [9]. For mitigation against this type of attack it is recommended to whitelist hosts [9] that can be specified in host headers.

### B. What is Broken Authentication?

Broken authentication [2] is a term used for several vulnerabilities which allow attackers to exploit and impersonate web application users [10]. There are various methods in which attackers can gain user credentials or hijack user sessions to be able to impersonate those users such as weak or guessable user credentials, incorrectly stored credentials such as passwords that have not been hashed which can be extracted from other types of attacks such as SQL injections, session IDs exposed in URLs, session fixation attacks, fixed session IDs and especially passwords, session IDs and other credentials which are sent over unencrypted connections such as HTTP [11]. Mitigation against this kind of attack include hashing passwords, ensuring session IDs are not exposed in URLs [11], settings timeouts on sessions so sessions are recreated after a certain amount of time [11], ensuring that passwords are not sent over unencrypted connections [11], using password security such as minimum lengths allowed and password complexity [11], hiding usernames and password error messages in unsuccessful logins so users and passwords cannot be brute forced [11] and ensuring protection against brute force such as using lockout features after 5 attempts [11].

### C. What is Sensitive Data Exposure?

Sensitive data exposure [2] refers to web applications which do not protect information such as passwords, financial information or health data which can lead to cyber criminals misusing this information to gain unauthorized access to user accounts, commit fraudulent acts such as making online purchases with the stolen payment information or commit blackmail with sensitive data acquired [12]. Sensitive data exposure can cause financial loses, damage the reputation of corporations who had their information or assets exposed and prompts businesses to pay the expense of investigating the data breaches. Protection against such attacks depends on the country legal frameworks and industry, since ignoring them can lead to financially devastating results [13].

### D. What are XML External Entities?

XML external entity injection [2] attacks also known as XXE injections occur when attackers abuse Extensible Mark-up Language (XML) parsers in web servers by sending specially crafted malicious XML documents to the web servers which are processed and can lead to denial of service, remote code execution or server side request forgery [14].

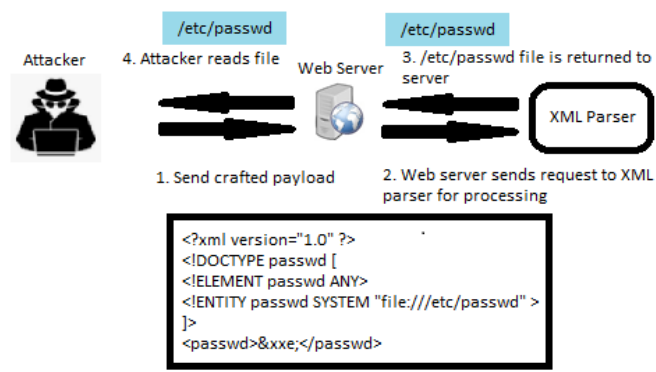

*Figure 2 - XML External Entity Attack Example*

XXE injection attacks have two types of attacks: in-band attacks which is when attackers craft a malicious XML file, submits online to be processed and receives the results in the same band [15] or in other words the attacker gets an instant response [16] and out-of-band attacks which occur when attackers craft the malicious XML files, submit to be processed but do not get a immediate response from the web server [16] which is also referred to as blind XXE injections. Figure 2 demonstrates how an attacker can abuse XML parsing to read internal system files. It is recommended to disable Document Type Definitions (DTD) to mitigate XXE injections [14].

### E. What is Broken Access Control?

Broken access control [2] consists of multiple possible attack vectors such as bypassing access control checks, editing other user's accounts, elevation of privileges, CORS misconfigurations which allow unauthorized access to restricted APIs, metadata manipulation through access control tokens like JSON Web Tokens (JWT) or access unauthorized web pages as a underprivileged user [17] which can lead to attackers control business functions or possibility of attackers obtaining all data [17]. It is recommended to use access control lists [17] and deny access to functionality by using server-side code where attackers cannot access or control metadata [17].

*F. What Are Security Misconfigurations?*

Security misconfigurations [2] refer to web applications that have been misconfigured in such a way that is leaving them exposed to security threats. They can include firewall misconfigurations [18], open administration ports [18] that expose the application to remote attacks or legacy applications that are trying to communicate with applications that do not exist anymore [18]. Ensuring configurations are going through a proper process of quality assurance and that such changes are thoroughly checked, tested and verified, reduces the attack surface from this type of vulnerability.

*G. What is Cross Site Scripting XSS?*

Cross site scripting [2] also known as XSS has many forms which lead to various outcomes based on the type of XSS being performed but typically occurs when attackers inject malicious scripts into the web application which then reveal sensitive information, internal services or disclose cookies of privileged users.

*Figure 3 - Common XSS Injection*

There are three types of XSS.
- Stored XSS. Stored XSS is when an attacker injects a payload into the webserver which is stored, so when another user requests to access the page the payload will trigger [19].
- Reflected XSS. Reflected XSS is when an attacker injects into POST data or the URL and is not stored but still reflects, by using this attackers can craft a malicious URL [19] to send this payload to users to and obtain sensitive information from them when triggering it such as their session cookies.
- DOM Based XSS is when an attacker injects their malicious payloads into the Document Object Model (DOM) and does not reflect in the HTML source and would only be able to trigger through the DOM console itself [20]. The vulnerability repercussions are the same as XSS.

Mitigation for XSS can be achieved by properly validating and normalizing data originating from untrusted sources [19].

*H. What is Insecure Deserialization?*

Insecure deserialization [2] attacks occur when applications try to transform malicious data, controlled by the attacker, into internal data structures controlled by the application [21]. By injecting specially crafted payloads, attackers can take control of variables, functions and internal application states. This often leads to remote code execution vulnerabilities [21] as well as exposure of the operating system of the web application.

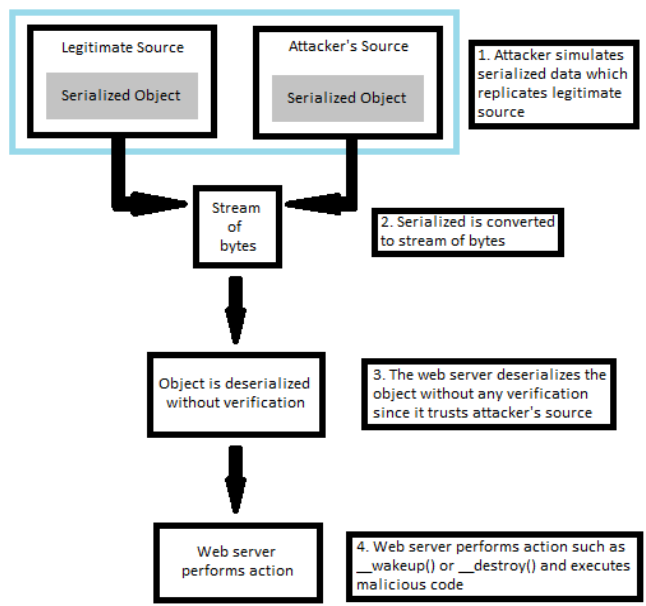

*Figure 4 - Deserialization Attack Example*

In code where serialization occurs an array of items are serialized into a stream of bytes which are then transported and processed by the backend of the website after being deserialized. When deserialization attacks occur, attackers replicate the serialized objects but injecting malicious code to be processed by the backend. Mitigation for this type of attack include using safer data-interchange formats such as JSON or YAML instead of native binary formats [21], using robust deserialization functions and libraries, including integrity checks [21] when processing serialized data and by limiting the scope and abilities of the serialized operations.

*I. What does Using Components With Known Vulnerabilities mean?*

Using components with known vulnerabilities [2] refers to using specific software or hardware with known vulnerabilities whether they have been discontinued or reach end of life [22]. Attackers are more likely to use common or known exploits to gain access to systems rather than discovering new vulnerabilities [22]. Protection against this vulnerability requires tracking of application dependencies, proper documentation [22], removal of unused dependencies, removal of dead code and inclusion of the dependencies into the application update policies, procedures and maintenance life cycle [22].

*J. What is Insufficient Logging & Monitoring?*

Insufficient logging and monitoring [2] refer to a lack of the proper logging mechanisms to assist in monitoring and detection of security incidents. This allows attackers to conduct their activities undetected which makes the task of incident detection and response from attacks [23] much harder. Logs are not only used for tracking attacker activity or detect errors and other anomalous activity that may take place on the application. Furthermore, many regulatory requirements depend on proper logging and monitoring mechanisms to be in place.

Some key mitigation points include implementation of logging facilities on key operations of the application, automated monitoring and detection mechanisms, ensuring proper storage policies are followed for the logs such as logs are secured and cannot be deleted [23].

## III. IMPACT ON BUSINESSES & USERS

The impact on business and users varies based on the type of attack and data sensitivity, but ultimately leaves businesses in financial losses. Some of the financial losses that may be incurred include: paying for incident investigation [24], customer compensation [24], damage control campaigns [25], operational down time [25] whilst the incident is investigated and resolved, possible legal actions [25] and fines if businesses violate regulatory compliance restrictions such as General Data Protection Regulations (GDPR) [24] which if violated can reach penalties of up to 4% of the annual turnover or 20 million Euros, whichever is greater [25]. From the customer's perspective trust is irreparable [25], once a breach has occurred, research showed that customers take their business to competitors, who seem to take security more seriously [25].

## IV. CONCLUSION

This report has covered the top 10 OWASP vulnerabilities are and how they impact businesses and web application users. The research conducted outlines the types of attacks, how to protect web servers from being affected and the impact on businesses and users who suffer from a breach. In conclusion monitoring, detecting and addressing vulnerabilities outlined can help businesses and end users in protecting against the most widely spread attacks and thus limiting their surface of attack and exposure. Moreover, basic knowledge of these types of attacks can help businesses to proactively secure against them and keep their assets and data protected.